\documentclass[showpacs,aps]{revtex4}
\usepackage{graphicx}
\usepackage{amsmath,amssymb,subfigure}
\usepackage{color}

\def\bc{\begin{center}}
\def\ec{\end{center}}

\def\beq{\begin{equation}}
\def\eeq{\end{equation}}

\begin{document}

\title{Coupling of two Dirac particles}
\author{Oleg L. Berman$^{1,2}$, Roman Ya. Kezerashvili$^{1,2}$, and Klaus
Ziegler$^{3}$}
\affiliation{\mbox{$^{1}$Physics
Department, New York City College
of Technology, The City University of New York,} \\
Brooklyn, NY 11201, USA \\
\mbox{$^{2}$The Graduate School and University Center, The
City University of New York,} \\
New York, NY 10016, USA \\
\mbox{$^3$   Institut f\"ur Physik, Universit\"at Augsburg\\
D-86135 Augsburg, Germany }}
\date{\today}

\begin{abstract}
A study of the formation of excitons as a problem of two Dirac particles in
a gapped graphene layer and in two gapped graphene layers separated by a
dielectric is presented. In the low-energy limit the separation of the
center-of-mass and relative motions is obtained. Analytical solutions for
the wave function and energy dispersion for both cases when electron and
hole interact via a Coulomb potential are found. It is shown that the energy
spectrum and the effective exciton mass are functions of the energy gaps as
well as interlayer separation in case of two layer gapped graphene.
\end{abstract}


\maketitle

$^{{}}$

\section{Introduction}

\label{intro}

\bigskip The exciton formed by an electron and a hole is one of the very
important objects in both experimental and theoretical physics of
semiconductor heterostructures and graphene. The study of the excitonic
system and its properties is necessary because of potential applications in
electronics and photonics, including, design of thresholdless lasers,
optical computing and quantum computing \cite{MacDonald 1990, Littlewood
2007, Butov 2008, Carusotto 2009, Shelykh 2010, Savona 2011, Snoke Solid
State physics, Amo Kavokin 2009}. From the theoretical point of view, an
exciton is a two-body system and to address the problem we have to solve a
two-body problem in semiconductor heterostructures or graphene. Let us
mention that the problem of the interaction between two particles is very
important for the deep analysis of the many-body physics in excitonic
system. The analysis of the spectrum of collective excitations necessary to
study such many-body phenomena in the excitonic system as Bose-Einstein
condensation and superfluidity~\cite{Snoke} requires the solution of
two-body problem for a single exciton. The temperature of
Kosterlitz-Thouless phase transition corresponding to the superfluidity
depends on the density of the superfluid component~\cite{Kosterlitz}. This
density of the superfluid component is defined by the two-particle Green
function of weakly interacting excitons. The problem to find the
two-particle Green function of the interacting excitons requires
one-particle Green function of non-interacting excitons, which is defined by
the wave function and energy dispersion of a single exciton~\cite{BLSC}. The
collective properties and superfluidity of excitons and polaritons in gapped
graphene have been studied in Refs.~\onlinecite{BKZ_1,BKZ_2}. Therefore, it
is of fundamental and practical interest to focus on solution of two-body
problem in semiconductor heterostructures and graphene. Finding solution of
two-body problem in gapped graphene layers is the subject of this Paper.

To describe the formation of an exciton in semiconductor heterostructures
like quantum wells the standard quantum mechanical approach is used based on
Schr\"{o}dinger equation. In this case the two-body problem with a scalar
action-at-a-distance inter-particle potentials is completely understood and
developed both in configuration or momentum three dimensional as well as two
dimensional (2D) spaces \cite{Landau, Liboff, Griffiths}. The notions of
absolute time and absolute space allow us to describe the two particles of
masses $m_{1\text{ }}$and $m_{2}$ with Euclidean position vectors $\mathbf{r}%
_{1\text{ }}$and $\mathbf{r}_{2},$ and momenta $\mathbf{p}_{1\text{ }}$and $%
\mathbf{p}_{2}$ in an inertial frame. It is well known~\cite{Landau} that in
the covariant mechanics, the most straightforward solution of the classical
problem with scalar interaction is obtained by a transformation from the
individual particle coordinates $\mathbf{r}_{1\text{ }}$, $\mathbf{r}_{2}$
and $\mathbf{p}_{1\text{ }}$, $\mathbf{p}_{2}$ to the covariant
center-of-mass and relative coordinates. Using canonical transformation

\begin{equation}
\mathbf{R}=\frac{m_{1}\mathbf{r}_{1}+m_{2}\mathbf{r}_{2}}{m}\ ,\text{ }%
\mathbf{r=r}_{1}-\mathbf{r}_{2}\text{ and }\mathbf{P}=\mathbf{p}_{1}+\mathbf{%
p}_{2},\text{ }\mathbf{p}=\mathbf{p}_{1}-\mathbf{p}_{2},
\end{equation}%
where $m=m_{1}+m_{2}$ is a total mass of the system, we can separate the
decoupled center-of-mass from the relative motion in configuration or
momentum space. In these coordinates, the nontrivial motion of the system of
two particles occurs entirely in the reduced problem of one body motion and
motion of the center-of-mass.\textit{\ }The total Hamiltonian of the system
can be presented as a sum of two parts. One part describes the motion of the
center-of-mass while the other two particles relative motion: \emph{H}=$%
\frac{\mathbf{P}^{2}}{2m}$+ $H_{r}$ with $H_{r}$ = $\frac{\mathbf{p}^{2}}{%
2\mu }+V(r),$ where $\mu =m_{1}m_{2}/m$ is two body's reduced mass. The
Hamiltonian $H_{r}$ governs the relative motion of two particles and, when
the Schr\"{o}dinger equation with this Hamiltonian have been solved, the
wave function of the relative motion of two particles is obtained.

Today graphene has been attracting a great deal of experimental and
theoretical attention because of unusual properties in its band structure~%
\cite{Castro_Neto_rmp,Das_Sarma_rmp}. Graphene is a two-dimensional layer of
carbon atoms, where the atoms form honeycomb lattice \cite{Novoselov1,
Zhang1}. In the low-energy limit the low energy excitations in graphene are
described not by the Schr\"{o}dinger equation but instead in graphene
electron and holes behave as relativistic massless particles described by a
Dirac-like equation for massless and chiral particles \cite{Semenoff,
DiVincenze,Guinea RMP 2009,Katsnelson,DasSarma}, which is known as Weyl
equation. Many of the unusual properties of graphene arise from the fact
that its quasiparticles are described by Dirac spinors. Since electron and
holes in graphene are governed by the Weyl equation they have an intrinsic
degree of freedom that resembles the spin degree of freedom in the original
Weyl equation. This degree of freedom is called pseudospin in order to
distinguish it from the spin and is described by the Pauli matrices. In
connection to the pseudospin, there is a good quantum number - the chirality
that is defined to be the projection of the 2-momentum operator on the
direction of the pseudospin. Clearly, the electrons will have positive
chirality and the holes will have negative chirality. Also the Fermi speed $%
v_{F}\sim 10^{6}$ m/s that is around 300 times slower than speed of light
replaces the speed of light in the original Weyl equation. The
positive-energy solutions of this 2D equation describes electrons, whereas
the negative-energy solutions describes holes.

A 2D exciton which is a bound state of an interacting electron and hole in
gapped graphene presents a two-body system and it is a fundamental and
practical interest to focus on solution of two-body problem in graphene.
Graphene consists of two equivalent carbon sublattices and
quantum-mechanical hopping between the sublattices leads to the formation of
two energy bands, and their intersection near the edges of the Brillouin
zone yields the conical energy spectrum. As a result, quasiparticles in
graphene exhibit a linear dispersion relation. The formation of excitons
requires the existence of the energy gap in electron and hole energy
spectrum. The formation of excitons in gapped graphene was studied in Refs.
\cite{Sabio, iyengar07, Berman 2008, BKZ_1}. The electronic ground state of
intrinsic graphene and bilayer graphene in the absence of the energy gap
using density functional theory within the local-density approximation and
Bethe-Salpeter equation was studied in Ref.~\cite{Yang1}. According to Ref.~%
\cite{Yang1}, no pure bound exciton was identified in intrinsic graphene and
bilayer graphene in the absence of the energy gap. Therefore, excitons in
graphene can be formed due to an energy gap opening in the electron and hole
spectra in the graphene layer. There are different mechanisms of electronic
excitations in graphene. The energy gap in graphene can be induced and
controlled by a magnetic field, doping, an electric field in biased
graphene, and hydrogenation \cite{iyengar07, Zhou, Lu, Haberer}.

By contrast quantum mechanics in the description of relativistic two-body
problem the situation is extremely more complicated and till now there is no
completely self-consistent theory for the separation of the center-of-mass
and relative motions even for the two-body case.\textbf{\ }This is due to
the following facts:

i) the particles locations and momenta are 4-vectors,

ii) the momenta are not independent but must satisfy mass-shell conditions,

iii) the inter-particle interaction potentials appear in the boosts as well
as in the energy generator in the instant form of dynamics. As a result of a
transformation to the center-of-mass system even a scalar
action-at-a-distance inter-particle interaction potential becomes dependent
on a coordinate and momentum.

iv) the structure of the Poincare' group implies that there is no definition
of relativistic 4-center of mass sharing all the properties of the
non-relativistic 3-center of mass \cite{Alba}.

The two body problem in graphene when the electron and hole are interacted
via a scalar action-at-a-distance inter-particle potential and are described
by Weyl's equation becomes even more complicated than in simple relativistic
case. Firstly, it is related to the fact that speed of light in the Weyl
equation for graphene is replaced by the the Fermi speed and the resultant
equation becomes non-covariant and canonical transformation implementing the
separation of the center-of-mass from the relative variables within
relativistic approach is invalided. Secondly, even the inter-particle
interaction depends only on the coordinate of the relative motion after to
the center-of-mass transformation it becomes dependent also on a momentum.
Lastly, the center-of-mass energy can not be separated from the relative
motion even though the interaction depends only on the coordinate of the
relative motion. This is caused by the chiral nature of Dirac electron in
graphene.

Our paper is organized in the following way. In Sec.~\ref{ham} we
present the Hamiltonian of the spatially separated electron and hole
in two different parallel graphene layers separated by a dielectric
in the presence of the band gap. In Sec. \ref{Sep} is given the
procedure of separation of the center-of-mass and relative motions
for two particles in graphene. In Sec.~\ref{2layers} we obtain the
energy spectrum and wave function of dipole exciton in two-layers
graphene and find the effective exciton mass. The energy spectrum
and wave function of exciton formed in a gapped graphene layer and
corresponding effective exciton mass are given in Sec.~\ref{1
layer}. Finally, the conclusions follow in Sec.~\ref{disc}.


\section{Exciton Hamiltonian}

\label{ham}

Let us consider two different parallel graphene layers with the interlayer
separation $D$ and assume that excitons in this system are formed by the
electrons located in the one graphene layer and, correspondingly, the holes
located in the other. In this system electrons and holes move in two
separate layers with honeycomb lattice structure. Since the motion of the
electron is restricted in one graphene layer and the motion of the hole is
restricted in the other graphene layer, we replace the coordinate vectors of
the electron and hole by their projections $\mathbf{r}_{1}$ and $\mathbf{r}%
_{2}$ on plane of one of the graphene sheet. These new in-plane coordinates $%
\mathbf{r}_{1}$ and $\mathbf{r}_{2}$ will be used everywhere below. Thus, we
reduced the restricted 3D two-body problem to the 2D two-body problem on the
graphene plane. Each honeycomb lattice is characterized by the coordinates $(%
\mathbf{r}_{j},1)$ on sublattice A and $(\mathbf{r}_{j},2)$ on sublattice B
with $j=1,2$ referring to the two sheets. Then the two-particle wave
function, describing two particles in different sheets, reads $\Psi (\mathbf{%
r}_{1},s_{1};\mathbf{r}_{2},s_{2})$, where $\mathbf{r}_{1}$ and $\mathbf{r}%
_{2}$ represent the coordinates of the electron and hole, correspondingly,
and $s_{1}$, $s_{2}$ are sublattice indices. This wave function can also be
understood as a four-component spinor, where the spinor components refer to
the four possible values of the sublattice indicies $s_{1},s_{2}$:
\begin{equation}
\Psi (\mathbf{r}_{1},s_{1};\mathbf{r}_{2},s_{2})=\left( {%
\begin{array}{c}
\phi _{aa}(\mathbf{r}_{1}\boldsymbol{,}\mathbf{r}_{2}) \\
\phi _{ab}(\mathbf{r}_{1},\mathbf{r}_{2}) \\
\phi _{ba}(\mathbf{r}_{1},\mathbf{r}_{2}) \\
\phi _{bb}(\mathbf{r}_{1},\mathbf{r}_{2})%
\end{array}%
}\right) \equiv \left( {%
\begin{array}{c}
\Psi _{a} \\
\Psi _{b}%
\end{array}%
}\right) ,\text{ where }\Psi _{a}=\left( {%
\begin{array}{c}
\phi _{aa} \\
\phi _{ab}%
\end{array}%
}\right) ,\ \ \ \Psi _{b}=\left( {%
\begin{array}{c}
\phi _{ba} \\
\phi _{bb}%
\end{array}%
}\right) .  \label{wave function1}
\end{equation}%
The two components mean that one particle is on sublattice a(b) and the
other particle is on sublattice a(b), correspondingly. In other words, the
spinor components are from the same tight-binding wave function at different
sites. Each graphene layer has an energy gap. Obviously the energy gaps in
graphene layers are independent and in the general case we can introduce two
non-equal gaps $\delta _{1}$ and $\delta _{2}$ for the first and the second
graphene layers, respectively.

The corresponding hopping matrix for two non-interacting particles,
including the energy gaps $\delta _{1}$ and $\delta _{2}$ on the first and
second layers, correspondingly, then reads
\begin{equation}
\mathcal{H}_{0}=\left(
\begin{array}{cccc}
-\delta _{1}+\delta _{2} & d_{2} & d_{1} & 0 \\
d_{2}^{\dag } & -\delta _{1}-\delta _{2} & 0 & d_{1} \\
d_{1}^{\dag } & 0 & \delta _{1}+\delta _{2} & d_{2} \\
0 & d_{1}^{\dag } & d_{2}^{\dag } & \delta _{1}-\delta _{2}%
\end{array}%
\right) \ .  \label{k17}
\end{equation}

In Eq. \eqref{k17} $d_{1}=\hbar v_{F}(-i\partial _{x_{1}}-\partial _{y_{1}})$%
, $d_{2}=\hbar v_{F}(-i\partial _{x_{2}}-\partial _{y_{2}})$ and the
corresponding hermitian conjugates are $d_{1}^{\dag }=\hbar v_{F}(-i\partial
_{x_{1}}+\partial _{y_{1}})$, $d_{2}^{\dag }=\hbar v_{F}(-i\partial
_{x_{2}}+\partial _{y_{2}})$, where $\partial _{x}=\partial /\partial x$ and
$\partial _{y}=\partial /\partial y,$ $x_{1}$, $y_{1}$ and $x_{2}$, $y_{2}$
are the coordinates of vectors $\mathbf{r}_{1}$ and $\mathbf{r}_{2}$,
correspondingly, and $v_{F}$ is the Fermi velocity of electrons in graphene.
This Hamiltonian allows us to write the eigenvalue equation for two
non-interacting particles as
\begin{equation}
\mathcal{H}_{0}\mathit{\Psi }_{0}=\epsilon _{0}\mathit{\Psi }_{0}\ ,
\label{k18}
\end{equation}%
%
%
%
%
%
%
%
%
%
%
which leads to the following eigenenergies:
\begin{equation}
\epsilon _{0}(k_{1},\delta _{1};k_{2},\delta _{2})=\pm \sqrt{\hbar
^{2}k_{1}^{2}+\hbar ^{2}k_{2}^{2}+\delta _{1}^{2}+\delta _{2}^{2}\pm 2\sqrt{%
(\hbar ^{2}k_{1}^{2}+\delta _{1}^{2})(\hbar ^{2}k_{2}^{2}+\delta _{2}^{2})}}%
=\pm \sqrt{\hbar ^{2}k_{1}^{2}+\delta _{1}^{2}}\pm \sqrt{\hbar
^{2}k_{2}^{2}+\delta _{2}^{2}}\ .  \label{5}
\end{equation}%
where $k_{1}$ and $k_{2}$ are momentum of each particle, correspondingly.
Eq.~\eqref{5} gives the energy spectrum for two non-interacting particles in
the presents of the non-equal gaps energies $\delta _{1}$ and $\delta _{2}$.
The energy dispersion is symmetrical with respect to the replacement of
particles $1$ and $2$. When there are no gaps, $\delta _{1}=0$ and $\delta
_{2}=0$, as it follows from \eqref{5} the energy dispersion is \ $\pm \hbar
k_{1}\pm \hbar k_{2}$.

Let's consider the electron and hole located in two graphene sheets with the
interlayer separation $D$, and interacting via the Coulomb potential $V(r),$
where $r$ is the projection of the distance between an electron and a hole
on the plane parallel to the graphene layers. Now the problem for the two
interacting particles located in different graphene layers with the broken
sublattice symmetry in each layer can be described by the Hamiltonian
\begin{equation}
\mathcal{H}=\left(
\begin{array}{cccc}
-\delta _{1}+\delta _{2}+V(r) & d_{2} & d_{1} & 0 \\
d_{2}^{\dag } & -\delta _{1}-\delta _{2}+V(r) & 0 & d_{1} \\
d_{1}^{\dag } & 0 & \delta _{1}+\delta _{2}+V(r) & d_{2} \\
0 & d_{1}^{\dag } & d_{2}^{\dag } & \delta _{1}-\delta _{2}+V(r)%
\end{array}%
\right) \ ,  \label{k20}
\end{equation}%
%
%
%
%
%
%
%
%
%
%
%
%
%
%
%
%
%
%
%
%
and the eigenvalue problem for Hamiltonian \eqref{k20} is
\begin{equation}
\mathcal{H}\Psi =\epsilon \Psi ,  \label{l}
\end{equation}%
where $\Psi $ are four-component eigenfunctions as given in Eq.%
\eqref{wave function1}.

The Hamiltonian \eqref{k20} describes two interacting particles located in
two graphene layers and satisfies the following conditions:

i) when both gaps vanish $\delta _{1}=0$ and $\delta _{2}=0$, as well as
two-body potential $V(r)=0$, the Hamiltonian describes two non-interacting
Dirac particles. It is important to mentioned that eigenenergies are
symmetrical with respect to the replacement of particle $1$ and $2$.

ii) when the interaction between particles vanishes, $V(r)=0,$ it describes
two independent particles, each located in the separate graphene layer,
having two independent gaps energies related to the broken sublattice
symmetry in each graphene sheet;

iii) when the gaps in each graphene layer vanish, $\delta _{1}=0$ and $%
\delta _{2}=0,$ the Hamiltonian describes two interacting particles in one
graphene layer interacting via Coulomb potential $V(r)\ $and is identical to
the Hamiltonian $(2)$ in Ref.~\cite{Sabio} representing the two-body problem
in one graphene layer if the band gap is absent;

iv) when the gaps $\delta _{1}=$ $\delta _{2}\equiv \delta ,$ the
Hamiltonian describes two interacting particles in one graphene layer
interacting via Coulomb potential $V(r)=e^{2}/4\pi \varepsilon
_{0}\varepsilon r,$ where $e$ is the charge of the electron and $\varepsilon
$ is the dielectric constant of the graphene layer.

v) when an electron and hole located in two different graphene sheets with
the interlayer separation $D$, they interacting via the potential $%
V(r)=-e^{2}/4\pi \varepsilon _{0}\varepsilon _{d}\sqrt{r^{2}+D^{2}},$ where $%
\varepsilon _{d}$ is the dielectric constant of the dielectric between two
graphene layers. Let us mention that for $\delta _{1}=\delta _{2}=0$ and $%
D=0 $ the Hamiltonian \eqref{k20} is also identical to the Hamiltonian $(2)$
in Ref.~\cite{Sabio}.


\section{Separation of the center-of-mass and relative motions}

\label{Sep}

In Hamiltonian~(\ref{k20}) the center-of-mass energy can not be separated
from the relative motion even though the interaction $V=V(r)$ depends only
on the coordinate of the relative motion. This is caused by the chiral
nature of Dirac electron in graphene. The similar conclusion was made for
the two-particle problem in graphene in Ref.~[\onlinecite{Sabio}].

Since the electron-hole Coulomb interaction depends only on the relative
coordinate, we introduce the new \textquotedblleft
center-of-mass\textquotedblright\ coordinates in the plane of a graphene
sheet:
\begin{gather}
\mathbf{R}=\alpha \mathbf{r}_{1}+\beta \mathbf{r}_{2}\ ,  \notag
\label{exp1} \\
\mathbf{r}=\mathbf{r}_{1}\boldsymbol{-}\mathbf{r}_{2}\ .
\end{gather}%
%
%
%
%
%
%
%
%
%
%
%
%
%
%
%
%
Here the coefficients $\alpha $ and $\beta $ are to be determined later.
Apparently we can use the analogy of the two-particle problem for Dirac
particles in gapped two-layer graphene with the center-of-mass coordinates
for the case of Schr\"{o}dinger equation. The coefficients $\alpha $ and $%
\beta $ will be found below from the condition of the separation of the
coordinates of the center-of-mass and relative motion in the Hamiltonian in
the one-dimensional \textquotedblleft scalar\textquotedblright\ equation
determining the corresponding component of the wave function.

To find the solution of \eqref{l} we make Anz\"{a}tze
\begin{equation}
\Psi _{j}(\mathbf{R}\boldsymbol{,}\mathbf{r})=\mathtt{e}^{i\mathbf{K}\cdot
\mathbf{R}}\psi _{j}(\mathbf{r})\ .
\end{equation}%
%
%
%
%
%
%
%
%
%
%
%
%
%
%
%
%
%
%
%
%
Let us introduce the following notations:
\begin{gather}
\mathcal{K}_{+}=\mathcal{K}\mathtt{e}^{i\Theta }=\mathcal{K}_{x}+i\mathcal{K}%
_{y}\ ,  \notag \\
\mathcal{K}_{-}=\mathcal{K}\mathtt{e}^{-i\Theta }=\mathcal{K}_{x}-i\mathcal{K%
}_{y}\ ,  \notag \\
\Theta =\tan ^{-1}\left( {\frac{\mathcal{K}_{y}}{\mathcal{K}_{x}}}\right) \ ,
\end{gather}%
%
%
%
%
%
%
%
%
%
%
%
%
%
%
%
%
%
%
%
%
and rewrite the Hamiltonian 
\eqref{k20} in terms of the representation of the coordinates $\mathbf{R}$
and $\mathbf{r}$ in a form of the $2\times 2$ matrix as
\begin{equation}
\mathcal{H}=\left( {\
\begin{array}{cc}
\mathcal{O}_{2}+V(r)\sigma _{0}-\delta _{1}\sigma _{0}+\delta _{2}\sigma _{3}
& \mathcal{O}_{1} \\
\mathcal{O}_{1}^{\dag } & \mathcal{O}_{2}+V(r)\sigma _{0}-\delta _{1}\sigma
_{0}+\delta _{2}\sigma _{3}%
\end{array}%
}\right) \ ,  \label{22ham}
\end{equation}%
%
%
%
%
%
%
%
%
%
%
%
%
%
%
%
%
%
%
%
%
where $\mathcal{O}_{1}$ and $\mathcal{O}_{2}$ are given by
\begin{gather}
\mathcal{O}_{1}=\hbar v_{F}\left( {\ \alpha \mathcal{K}_{-}-i\partial
_{x}-\partial _{y}}\right) \sigma _{0},  \label{oo} \\
\mathcal{O}_{2}=\hbar v_{F}\left( {\
\begin{array}{cc}
0 & \beta \mathcal{K}_{-}+i\partial _{x}+\partial _{y} \\
\beta \mathcal{K}_{+}+i\partial _{x}-\partial _{y} & 0%
\end{array}%
}\right)  \label{000}
\end{gather}%
where $x$ and $y$ are the components of vector $\mathbf{r}$, $\sigma _{j}$
are the Pauli matrices, $\sigma _{0}$ is the $2\times 2$ unit matrix.
Analysis of the operators \eqref{oo} and \eqref{000} shows that the
coordinates of the center-of-mass and relative motion can be separated in a
certain approximation.

For the Hamiltonian \eqref{22ham} the eigenvalue problem $\mathcal{H}\Psi
=\epsilon \Psi $ results in the following equations:
\begin{eqnarray}
\left( {\mathcal{O}_{2}+V(r)\sigma _{0}-\delta _{1}\sigma _{0}+\delta
_{2}\sigma _{3}}\right) \Psi _{a}+\mathcal{O}_{1}\Psi _{b} &=&\epsilon
\sigma _{0}\Psi _{a}  \notag \\
\mathcal{O}_{1}^{\dag }\Psi _{a}+\left( {\mathcal{O}_{2}+V(r)\sigma
_{0}-\delta _{1}\sigma _{0}+\delta _{2}\sigma _{3}}\right) \Psi _{b}
&=&\epsilon \sigma _{0}\Psi _{b}\ .  \label{221}
\end{eqnarray}

%
%
%
%
From Eq.~\eqref{221} we have:
\begin{equation}
\Psi _{b}=\left( {\epsilon \sigma _{0}-\mathcal{O}_{2}-V(r)\sigma
_{0}+\delta _{1}\sigma _{0}-\delta _{2}\sigma _{3}}\right) ^{-1}\mathcal{O}%
_{1}^{\dag }\Psi _{a}\ .  \label{a2}
\end{equation}%
%
%
%
%
%
%
%
%
%
%
%
%
%
%
%
%
%
%
%
%
Assuming the interaction potential and both relative and center-of-mass
kinetic energies are small compared to the gaps $\delta _{1}$ and $\delta
_{2}$ we use the following approximation:
\begin{equation}
\left( {\epsilon \sigma _{0}-\mathcal{O}_{2}-V(r)\sigma _{0}+\delta
_{1}\sigma _{0}-\delta _{2}\sigma _{3}}\right) ^{-1}\backsimeq \frac{1}{%
\epsilon \sigma _{0}+\delta _{1}\sigma _{0}-\delta _{2}\sigma _{3}}\ .
\end{equation}%
%
%
%
%
%
%
%
%
%
%
%
%
%
%
%
%
%
%
%
%
Using the fact that the operator $\mathcal{O}_{1}^{\dag }\mathcal{O}_{1}$ is
purely hermitian, applying Eq.~\eqref{221} and
\begin{equation}
\mathcal{O}_{1}^{\dag }\mathcal{O}_{1}=\hbar ^{2}v_{F}^{2}\left( {\alpha ^{2}%
\mathcal{K}^{2}-\nabla _{\mathbf{r}}^{2}-2i\alpha (\mathcal{K}_{x}\partial
_{y}+\mathcal{K}_{y}\partial _{x})}\right) \sigma _{0}\ ,
\end{equation}%
%
%
%
%
%
%
%
%
%
%
%
%
%
%
%
%
%
%
%
%
we obtain:
\begin{equation}
\left( {\mathcal{O}_{2}+V(r)\sigma _{0}-\delta _{1}\sigma _{0}+\delta
_{2}\sigma _{3}}\right) \Psi _{a}+\hbar ^{2}v_{F}^{2}\frac{\left( {\alpha
^{2}\mathcal{K}^{2}-\nabla _{\mathbf{r}}^{2}-2i\alpha (\mathcal{K}%
_{x}\partial _{x}+\mathcal{K}_{y}\partial _{y})}\right) }{\epsilon \sigma
_{0}+\delta _{1}\sigma _{0}-\delta _{2}\sigma _{3}}\Psi _{a}=\epsilon \sigma
_{0}\Psi _{a}\ .  \label{a5}
\end{equation}%
%
%
%
%
%
%
%
%
%
%
%
%
%
%
%
%
%
%
%
%

Now using Eq. \eqref{wave function1} we rewrite Eq.~\eqref{a5} for
the individual spinor components in the following form:
\begin{eqnarray}
&&\left( {\ -\delta _{1}+\delta _{2}+V(r)+\hbar ^{2}v_{F}^{2}\frac{\alpha
^{2}\mathcal{K}^{2}-\nabla _{\mathbf{r}}^{2}-2i\hbar v_{F}\alpha (\mathcal{K}%
_{x}\partial _{x}+\mathcal{K}_{y}\partial _{y})}{\epsilon -\delta
_{1}-\delta _{2}}}\right) \phi _{aa}+  \notag \\
&&\hbar v_{F}\left( {\beta \mathcal{K}_{-}+i\partial _{x}+\partial _{y}}%
\right) \phi _{ab}=\epsilon \phi _{aa}\ ,  \label{231}
\end{eqnarray}%
\begin{eqnarray}
&&\hbar v_{F}\left( {\ \beta \mathcal{K}_{+}+i\partial _{x}-\partial _{y}}%
\right) \phi _{aa}+  \notag \\
&&\left( {-\delta _{1}-\delta _{2}+V(r)+\hbar ^{2}v_{F}^{2}\frac{\alpha ^{2}%
\mathcal{K}^{2}-\nabla _{\mathbf{r}}^{2}-2i\alpha (\mathcal{K}_{x}\partial
_{x}+\mathcal{K}_{y}\partial _{y})}{\epsilon -\delta _{1}+\delta _{2}}}%
\right) \phi _{ab}=\epsilon \phi _{ab}\ .  \label{a7}
\end{eqnarray}%
%
%
%
%
%
%
%
%
%
%
%
%
%
%
%
%
%
%
%
%
We solve Eq.~\eqref{a7} with respect to $\phi _{ab}$:
\begin{equation}
\phi _{ab}=\left[ {\epsilon +\delta _{1}+\delta _{2}-V(r)-\hbar ^{2}v_{F}^{2}%
\frac{\alpha ^{2}\mathcal{K}^{2}-\nabla _{\mathbf{r}}^{2}-2i\alpha (\mathcal{%
K}_{x}\partial _{x}+\mathcal{K}_{y}\partial _{y})}{\epsilon -\delta
_{1}+\delta _{2}}}\right] ^{-1}\left( {\ \beta \mathcal{K}_{+}+i\partial
_{x}-\partial _{y}}\right) \hbar v_{F}\phi _{aa}\ .  \label{ab}
\end{equation}%
%
%
%
%
%
%
%
%
%
%
%
%
%
%
%
%
%
%
%
%
Substituting $\phi _{ab}$ from Eq.~(\ref{ab}) into Eq.~(\ref{231}), we
obtain:
\begin{eqnarray}
&&\left( {\ -\delta _{1}+\delta _{2}+V(r)+\hbar ^{2}v_{F}^{2}\frac{\alpha
^{2}\mathcal{K}^{2}-\nabla _{\mathbf{r}}^{2}-2i\alpha (\mathcal{K}%
_{x}\partial _{x}+\mathcal{K}_{y}\partial _{y})}{\epsilon -\delta
_{1}-\delta _{2}}}\right) \phi _{aa}+  \notag \\
&+&\hbar ^{2}v_{F}^{2}\left( {\beta \mathcal{K}_{-}+i\partial _{x}+\partial
_{y}}\right) \left[ {\epsilon +\delta _{1}+\delta _{2}-V(r)-\hbar
^{2}v_{F}^{2}\frac{\alpha ^{2}\mathcal{K}^{2}-\nabla _{\mathbf{r}%
}^{2}-2i\alpha (\mathcal{K}_{x}\partial _{x}+\mathcal{K}_{y}\partial _{y})}{%
\epsilon -\delta _{1}+\delta _{2}}}\right] ^{-1}  \notag \\
&\times &\left( {\ \beta \mathcal{K}_{+}+i\partial _{x}-\partial _{y}}%
\right) =\epsilon \phi _{aa}\ .  \label{a10}
\end{eqnarray}%
%
%
%
%
%
%
%
%
%
%
%
%
%
%
%
%
%
%
%
%
Assuming again that the interaction potential and both relative and
center-of-mass kinetic energies are small compared to the gaps $\delta _{1}$
and $\delta _{2}$ we apply to Eq.~(\ref{a10}) the following approximation:
\begin{equation}
\left[ {\epsilon +\delta _{1}+\delta _{2}-V(r)-\hbar ^{2}v_{F}^{2}\frac{%
\alpha ^{2}\mathcal{K}^{2}-\nabla _{\mathbf{r}}^{2}-2i\alpha (\mathcal{K}%
_{x}\partial _{x}+\mathcal{K}_{y}\partial _{y})}{\epsilon -\delta
_{1}+\delta _{2}}}\right] ^{-1}=\frac{1}{\epsilon +\delta _{1}+\delta _{2}}\
.  \label{a11}
\end{equation}%
%
%
%
%
%
%
%
%
%
%
%
%
%
%
%
%
%
%
%
%
Applying the approximation given by Eq.~(\ref{a11}) to Eq.~(\ref{a10}), we
get from Eq.~(\ref{a10}) the eigenvalue equation spinor component $\phi
_{aa} $ in the form:

\begin{eqnarray}  \label{fin1}
&& \left({-\delta _{1}+\delta _{2}+V(r)+\hbar ^{2}v_{F}^{2}\frac{\alpha ^{2}%
\mathcal{K}^{2}-\nabla _{\mathbf{r}}^{2}-2i\alpha (\mathcal{K}_{x}\partial
_{x}+\mathcal{K}_{y}\partial _{y})}{\epsilon -\delta _{1}-\delta _{2}}}
\right.  \notag \\
&& \left.+ \hbar ^{2}v_{F}^{2}\frac{\beta ^{2}\mathcal{K}^{2}-\nabla _{%
\mathbf{r}}^{2}-2i\beta (\mathcal{K}_{x}\partial _{x}+\mathcal{K}%
_{y}\partial _{y})}{\epsilon -\delta _{1}-\delta _{2}}\right)\phi
_{aa}=\epsilon \phi _{aa}\ .
\end{eqnarray}

Choosing the values for the coefficients $\alpha $ and $\beta $ to separate
the coordinates of the center-of-mass (the wave vector $\mathbf{\mathcal{K}}$%
) and relative motion $\mathbf{r}$ in Eq.~(\ref{fin1}), we have
\begin{eqnarray}
\alpha &=&\frac{\epsilon -\delta _{1}-\delta _{2}}{2\epsilon }\ ,  \notag \\
\beta &=&\frac{\epsilon +\delta _{1}+\delta _{2}}{2\epsilon }\ .  \label{a13}
\end{eqnarray}%
%
%
%
%
%
%
%
%
%
%
%
%
%
%
%
%
%
%
%
%
After substitution of Eq.~(\ref{a13}) into Eq.~(\ref{fin1}) we can
obtain the component $\phi _{aa}$ of the spinor \eqref{wave
function1} as a solution of one-dimensional second order
differential equation:
\begin{equation}
\left( {\ \frac{(\hbar v_{F}\mathcal{K})^{2}}{2\epsilon }+V(r)-\frac{%
\epsilon (\hbar v_{F})^{2}\nabla _{\mathbf{r}}^{2}}{2\left( {\epsilon
^{2}-(\delta _{1}+\delta _{2})^{2}}\right) }}\right) \phi _{aa}=\left[ {%
\epsilon +\delta _{1}-\delta _{2}}\right] \phi _{aa}\ .  \label{fin11}
\end{equation}%
The other components of \eqref{wave function1} can be found as:
\begin{equation}
\Psi _{b}=-(\epsilon \sigma _{0}+(\partial _{x_{2}}\sigma _{1}-i\partial
_{y_{2}}\sigma _{2})-\delta _{1}\sigma _{0}-\delta _{2}\sigma
_{3}-V(r)\sigma _{0})^{-1}iD_{1}^{\dagger }\Psi _{a}  \label{eq:8}
\end{equation}%
with Pauli matrices $\sigma _{j}$ and $2\times 2$ unit matrix $\sigma _{0}$.
Moreover, we have
\begin{equation}
\phi _{ab}=\left[ \epsilon +\delta _{1}+\delta _{2}-V(r)+\frac{1}{\epsilon
-\delta _{1}+\delta _{2}}(\partial _{x_{1}}^{2}+\partial _{y_{1}}^{2})\right]
^{-1}(i\partial _{x_{2}}-\partial _{y_{2}})\phi _{aa}\ .  \label{eq:7}
\end{equation}

\bigskip

\section{\protect\bigskip Two body problem in two gapped graphene layers}

\label{2layers}

Let us consider an electron and hole located in two different parallel
graphene layers and interacting via the potential $V(r)=-e^{2}/4\pi
\varepsilon _{0}\varepsilon _{d}\sqrt{r^{2}+D^{2}}.$ Substituting this
potential into (\ref{fin11}) we obtain the second order differential
equation for the component $\phi _{aa}.$ This equation cannot be solved
analytically. However, assuming $r\ll D$ we can approximate $V(r)$ by the
first two terms of Taylor series and substituting $V(r)=-V_{0}+\gamma r^{2},$
where $V_{0}=e^{2}/4\pi \varepsilon _{0}\varepsilon _{d}D$ and $\gamma
=e^{2}/8\pi \varepsilon _{0}\varepsilon _{d}D^{3}$ for the interaction
potential into Eq.~(\ref{fin11}), we obtain
\begin{equation}
\left( {-\frac{2\epsilon (\hbar v_{F})^{2}\nabla _{\mathbf{r}}^{2}}{{%
\epsilon ^{2}-(\delta _{1}+\delta _{2})^{2}}}+\gamma r^{2}}\right) \phi
_{aa}=\left[ {\epsilon +\delta _{1}-\delta _{2}+V_{0}-\frac{(\hbar v_{F}%
\mathcal{K})^{2}}{2\epsilon }}\right] \phi _{aa}\ .  \label{2simpl}
\end{equation}%
%
%
%
%
%
%
%
%
%
%
%
%
%
%
%
%
%
%
%
%
The last equation can be rewritten in the form of the two-dimensional
isotropic harmonic oscillator:
\begin{equation}
\left( {-\mathcal{F}_{1}(\epsilon )\nabla _{\mathbf{r}}^{2}+\gamma r^{2}}%
\right) \phi _{aa}=\mathcal{F}_{0}(\epsilon )\phi _{aa}\ ,  \label{harm}
\end{equation}%
%
%
%
%
%
%
%
%
%
%
%
%
%
%
%
%
%
%
%
%
where
\begin{gather}
\mathcal{F}_{1}=\frac{2\epsilon (\hbar v_{F})^{2}}{{\epsilon ^{2}-(\delta
_{1}+\delta _{2})^{2}}}\ ,  \notag \\
\mathcal{F}_{0}=\epsilon +\delta _{1}-\delta _{2}+V_{0}-\frac{(\hbar v_{F}%
\mathcal{K})^{2}}{2\epsilon }\ .
\end{gather}%
%
%
%
%
%
%
%
%
%
%
%
%
%
%
%
%
%
%
%
%
The eigenfunction and eigenenergy for two-dimensional isotropic parabolic
well were first determined by Fock in 1928~\cite{Fock}, later in Ref. [%
\onlinecite{Darwin}] , and was studied in detail in Ref. \cite{Dingle}. The
single-particle eigenfunction for the two-dimensional oscillator widely used
for the description of quantum dot \cite{Maksym}. The solution of Eq.~(\ref%
{harm}) is well known (see, for example, Ref.~\cite{arfken85}) and is given
by
\begin{equation*}
\frac{\mathcal{F}_{0}(\epsilon )}{\mathcal{F}_{1}(\epsilon )}=2N\sqrt{\frac{%
\gamma }{\mathcal{F}_{1}(\epsilon )}}\ ,
\end{equation*}%
%
%
%
%
%
%
%
%
%
%
%
%
%
%
%
%
%
%
%
%
where $N=2\tilde{N}+|L|+1$, and $\tilde{N}=\mathrm{min}(\widetilde{n},%
\widetilde{n}^{\prime })$, $L=\widetilde{n}-\widetilde{n}^{\prime }$, $%
\widetilde{n},$ $\widetilde{n}^{\prime }=0,1,2,3,\ldots $ are the quantum
numbers of the 2D harmonic oscillator. The corresponding 2D wave function in
terms of associated Laguerre polynomials can be written as
\begin{equation}
\phi _{aa_{\tilde{N}\text{ }L}}(r)=\frac{\tilde{N}!}{a^{|L|+1}\sqrt{%
\widetilde{n}!\widetilde{n}^{\prime }!}}\mathrm{sgn}%
(L)^{L}r^{|L|-1/2}e^{-r^{2}/(4a^{2})}\times L_{\tilde{N}%
}^{|L|}(r^{2}/(2a^{2}))\frac{e^{-iL\phi }}{(2\pi )^{1/2}}\ ,  \label{rk14}
\end{equation}%
where $\phi $ is the polar angle, $L_{k}^{p}(x)$ are the associated Laguerre
polynomials and $a=\left( \sqrt{F_{1}(\epsilon )}/\left( 2\sqrt{\gamma }%
\right) \right) ^{1/2}$.

\begin{figure}[ht]
\centering
\subfigure[]{
\includegraphics[width=0.48\textwidth]{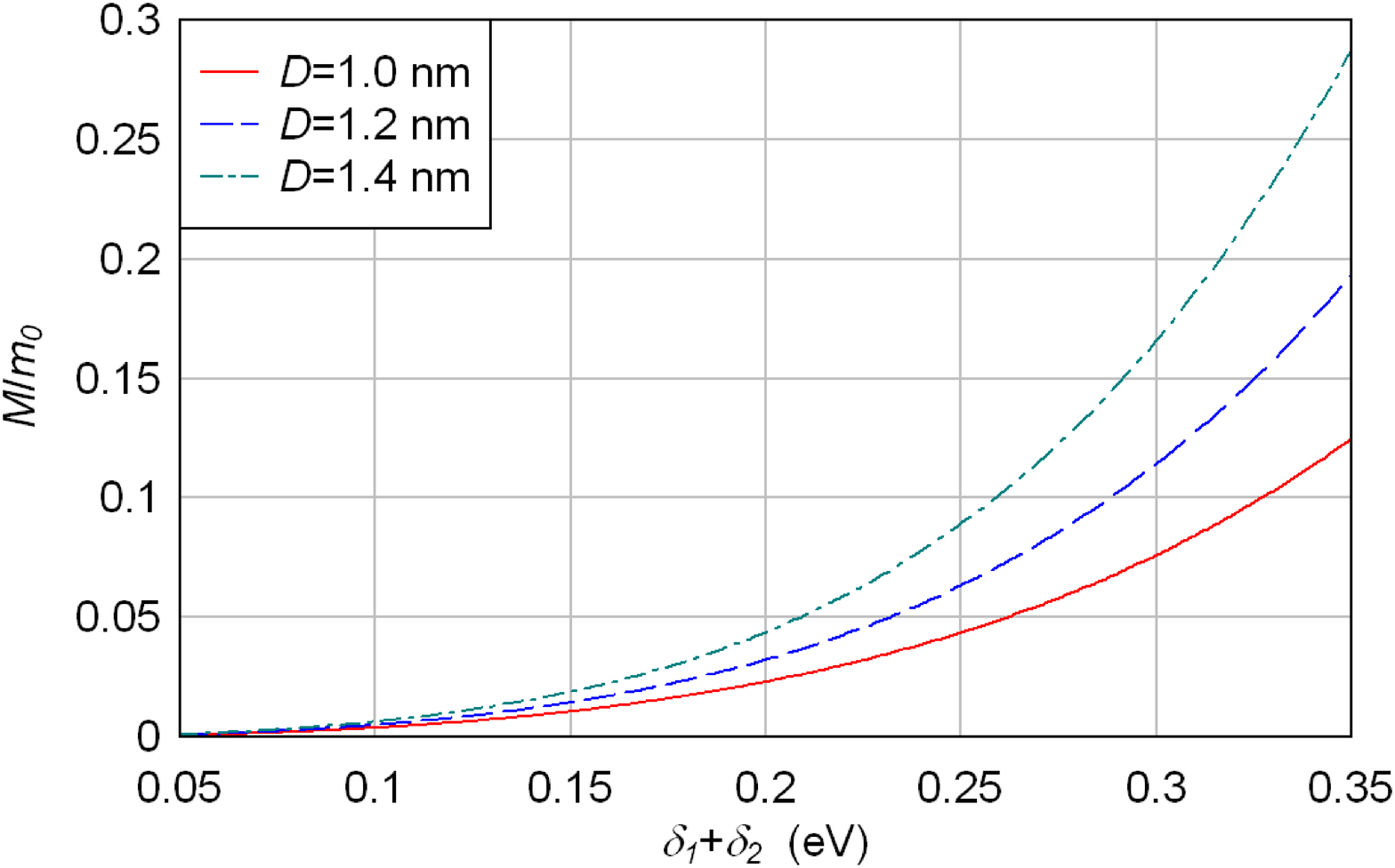}
} \subfigure[]{
\includegraphics[width=0.48\textwidth]{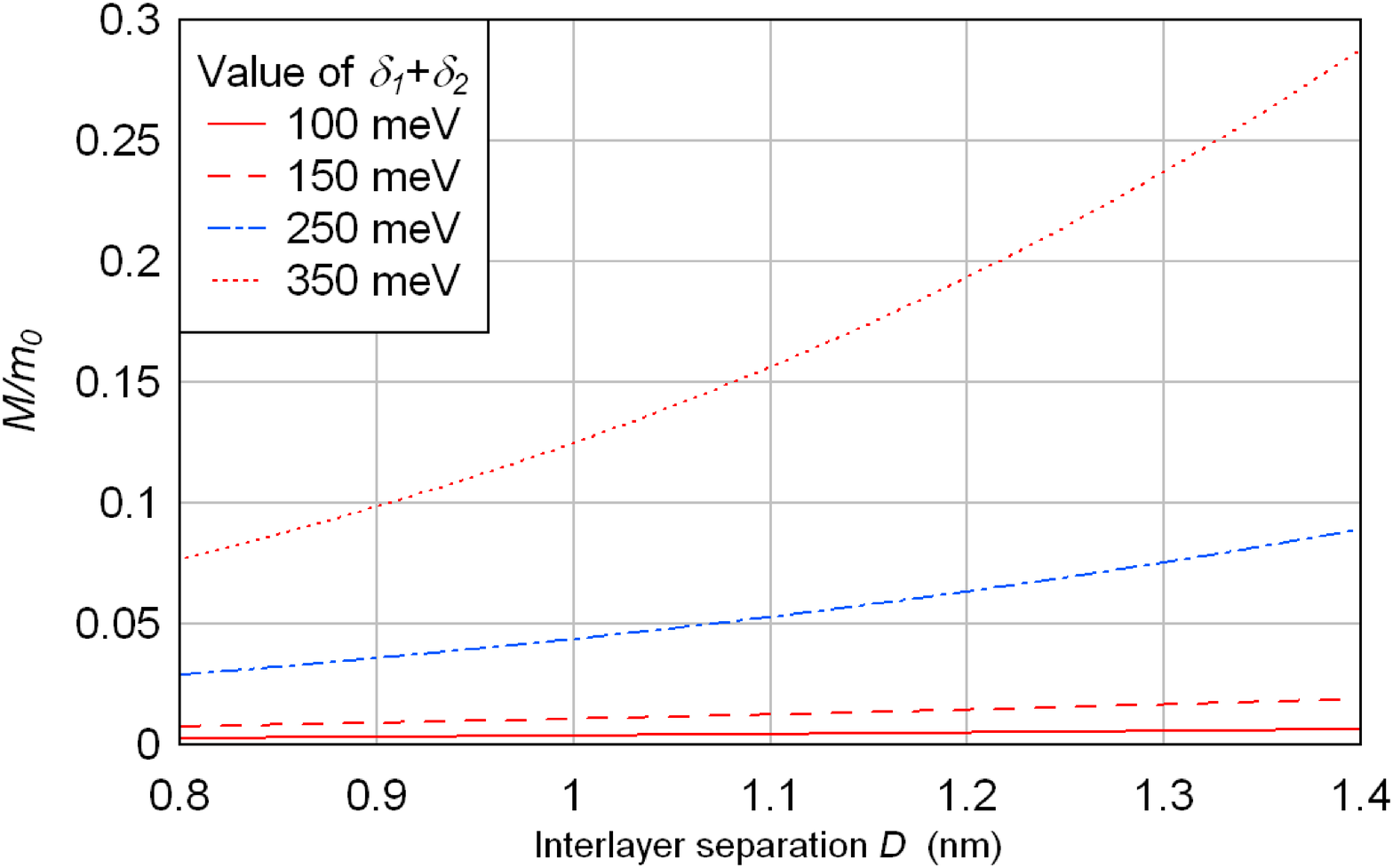}
}
\caption{Excitons in two graphene layers separated by the dielectric GaAs.
The effective exciton mass $M$ in the units of free electron mass $m_0$ (a)
as a function of the total energy gap for the different graphene interlayer
separations, (b) as the function of on interlayer separation for different
values of the total energy gap.}
\label{F1}
\end{figure}

After some straightforward but lengthy calculations and the expansion up to
second order in $\mathcal{K}$ we obtain the following expression for the
energy in quadratic order with respect to $\mathcal{K}$
\begin{equation}
\epsilon =-V_{0}+\sqrt{\mu ^{2}+\frac{C_{1}}{\mu }}+\frac{1}{2\mu ^{4}}\frac{%
C_{1}}{\sqrt{1+\frac{C_{1}}{\mu ^{3}}}}(\hbar v_{F}\mathcal{K})^{2}\ ,
\label{energy5}
\end{equation}%
%
%
%
%
%
%
%
%
%
%
%
%
%
%
%
%
%
where $\mu =\delta _{1}+\delta _{2}$ and $C_{1}=2\gamma N^{2}(\hbar
v_{F})^{2}$. Thus, from \eqref{energy5} we can conclude that the effective
exciton mass $M$ is given as a function of total energy gap $\delta
_{1}+\delta _{2}$ as
\begin{equation}
M=\frac{\mu ^{4}}{v_{F}^{2}C_{1}}\sqrt{1+\frac{C_{1}}{\mu ^{3}}}\ .
\label{mass}
\end{equation}%
%
%
%
%
%
%
%
%
%
%
%
%
%
%
%
%
%
The dependence of the effective exciton mass $M$ defined by Eq. \eqref{mass}
on the total energy gap $\delta _{1}+\delta _{2}$ and the interlayer
separation $D$ for two graphene layers separated by the dielectric GaAs is
shown in Fig.~\ref{F1}. Firstly, according to Fig.~\ref{F1}, the effective
exciton mass $M$ increases when the total energy gap $\delta _{1}+\delta
_{2} $ and the interlayer separation $D$ increase. Secondly, the mass
increases much faster for the bigger value of the interlayer separation and
bigger value of the energy gap. Let's mentioned that the dependence of the
effective exciton mass $M$ on the interlayer separation $D$ is caused by the
quasi-relativistic Dirac Hamiltonian of the gapped electrons and holes in
graphene layers. However, for the excitons in couple quantum wells (CQW) the
effective exciton mass does not depend on the interlayer separation, because
the electrons and holes in CQW's are described by a Schr\"{o}dinger
Hamiltonian, while excitons in two graphene layers are described by the
Dirac-like Hamiltonian \eqref{k20}.

\begin{figure}[ht]
\centering
\subfigure[]{
\includegraphics[width=0.48\textwidth]{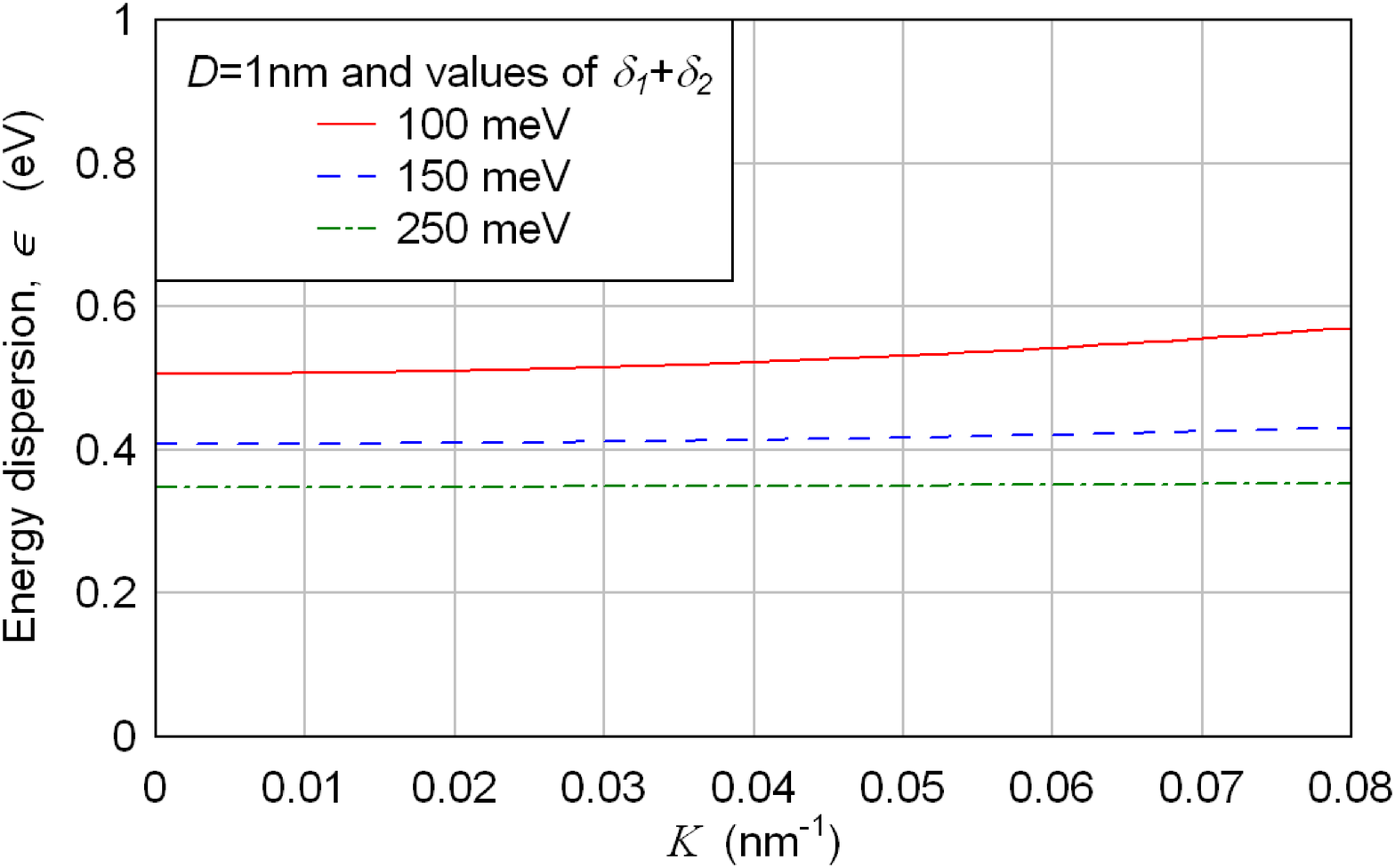}
} \subfigure[]{
\includegraphics[width=0.48\textwidth]{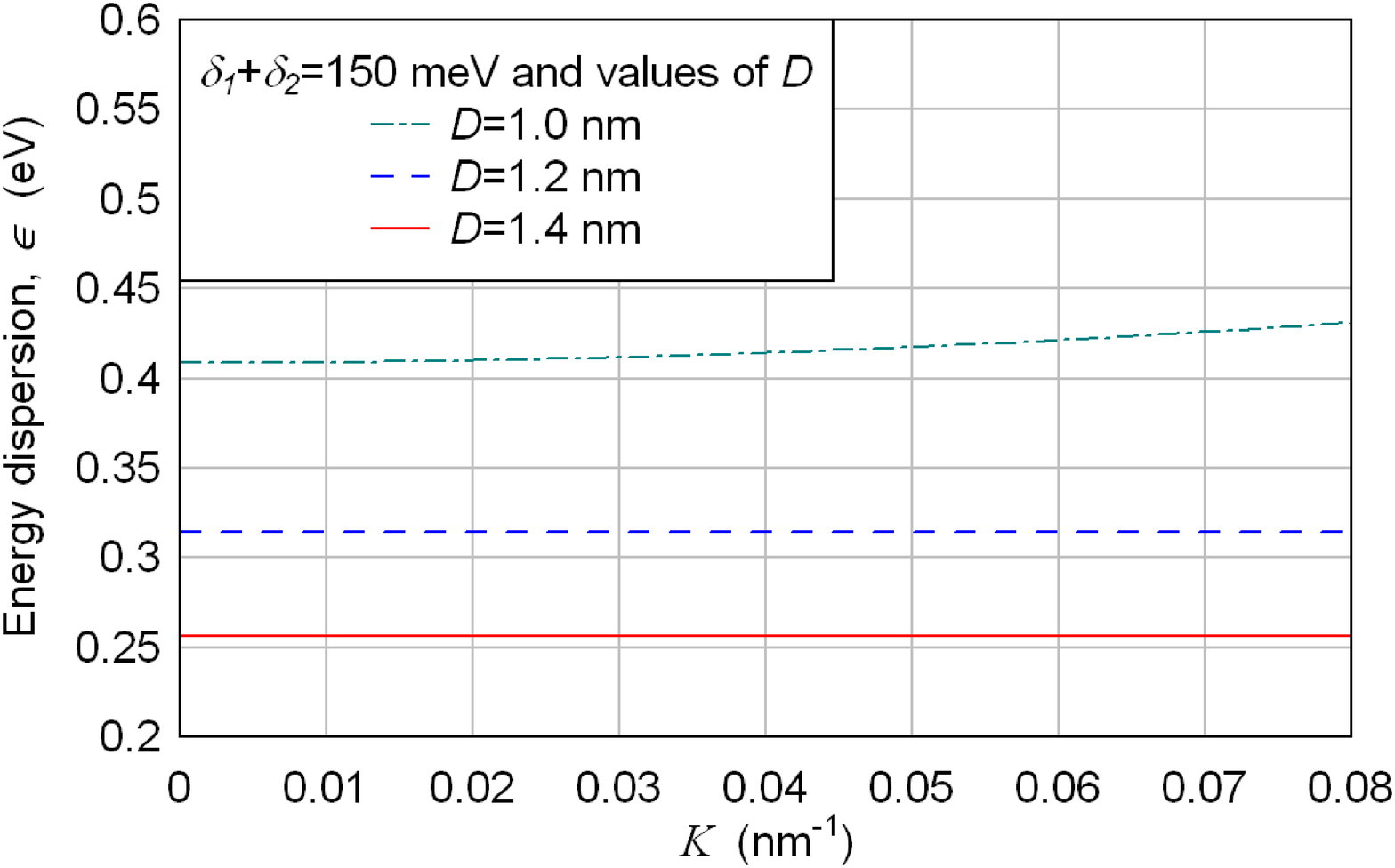}
} \subfigure[]{
\includegraphics[width=0.48\textwidth]{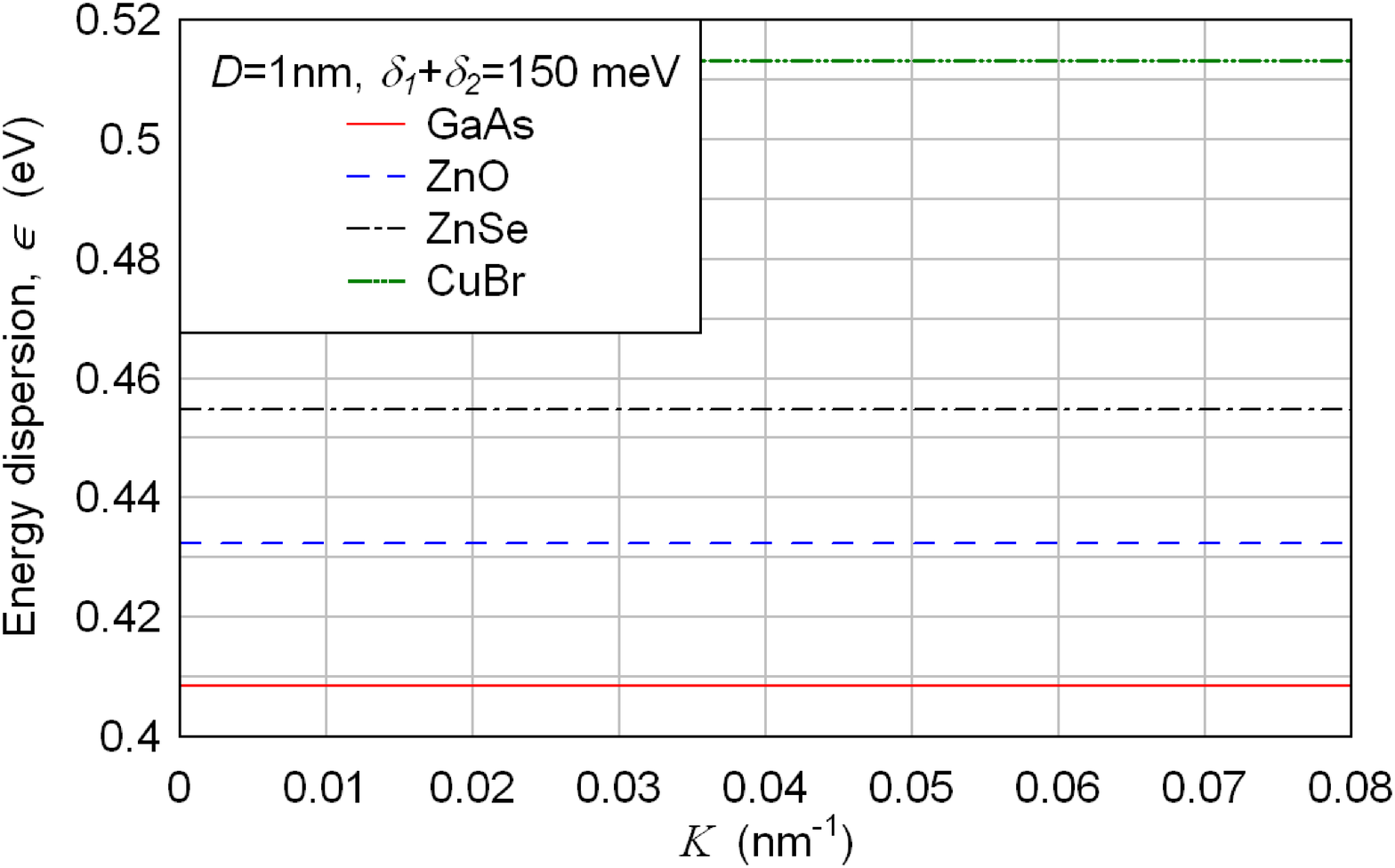}
}
\caption{Excitons in two graphene layers separated by the dielectric. The
energy dispersion of exciton (a) for the fixed total energy gap and the
different graphene interlayer separations, (b) for the fixed interlayer
separation for different values of the total energy gap, and (c) for the
fixed total energy gap, graphene interlayer separations and the different
dielectrics between graphene layers.}
\label{F2}
\end{figure}

In Fig.~\ref{F2} is shown the energy dispersion of exciton for different
values of the total energy gap and interlayer separation and for the
different dielectrics between the graphene layers. Results are presented for
the parabolic approximation for the energy dispersion assuming the
low-energy limit. When $\mathcal{K\sim }$0.08 nm$^{-1}$ the parabolic
approximation \eqref{energy5} gives about 2\% difference with respect to the
exact numerical solution and this percentage decreases when $\mathcal{K}$
deceases. The analysis of the results presented in Fig. \ref{F2}a shows that
the energy dispersion decreases when the total energy gap increases. The
same behavior can be observed for the dependence of the energy dispersion on
interlayer separation: for given $\mathcal{K}$ \ when interlayer separation
increases the energy dispersion decreases (Fig. \ref{F2}b). In Fig.~\ref{F2}%
c is given the energy dispersion of exciton for the different dielectric
placed between two graphene layers. As it seen, there is small increase for
the energy dispersion: for a smaller value of the dielectric constant the
energy dispersion of exciton becomes bigger.

\section{Two body problem in a gapped graphene layer}

\label{1 layer}

Now we consider an electron and a hole located in a single gapped graphene
layer with the energy gap parameter $\delta $ and exciton formed by the
electron and hole located in this graphene layer. Putting the gaps $\delta
_{1}=$ $\delta _{2}\equiv \delta $ in \eqref{k20}, we obtain the Hamiltonian
that describes two interacting particles in one graphene layer interacting
via Coulomb potential $V(r)=\frac{e^{2}}{4\pi \varepsilon _{0}\varepsilon r}%
. $ Using this potential we rewrite Eq.~(\ref{fin11}) in the form
\begin{equation}
\left( {-\frac{2\epsilon (\hbar v_{F})^{2}\nabla _{\mathbf{r}}^{2}}{{%
\epsilon ^{2}-4\delta ^{2}}}-\frac{e^{2}}{4\pi \varepsilon _{0}\varepsilon r}%
}\right) \phi _{aa}=\left[ {\epsilon -\frac{(\hbar v_{F}\mathcal{K})^{2}}{%
2\epsilon }}\right] \phi _{aa}\ .  \label{3simpl}
\end{equation}%
%
%
%
%
%
%
%
%
%
%
%
%
%
%
%
%
%
%
%
%

Eq.~(\ref{3simpl}) can be rewritten in the form of the two-dimensional
hydrogen atom:
\begin{equation}
\left( -\mathcal{F}_{1}(\epsilon )\nabla _{r}^{2}-\frac{e^{2}}{4\pi
\varepsilon _{0}\varepsilon r}\right) \phi _{aa}=\mathcal{F}_{0}(\epsilon
)\phi _{aa}\ ,
\end{equation}%
%
%
%
%
%
%
%
%
%
%
%
%
%
%
%
%
%
%
%
%
%
%
%
where
\begin{equation}
\mathcal{F}_{1}=\frac{2\epsilon (\hbar v_{F})^{2}}{{\epsilon ^{2}-4\delta
^{2}}}\ ,\ \ \ \mathcal{F}_{0}={\epsilon }-\frac{(\hbar v_{F}\mathcal{K})^{2}%
}{2\epsilon }\ .
\end{equation}%
%
%
%
%
%
%
%
%
%
%
%
%
%
%
%
%
%
%
%
%
%
%
%
The solution of the two-dimensional hydrogen atom equation (\ref{harm}) is
well known \cite{Zaslow, Problem1, Yang} and is given by
\begin{equation}
\mathcal{F}_{0}(\epsilon )=-\frac{e^{4}}{4\mathcal{F}_{1}(\epsilon
)\varepsilon ^{2}\left( n-1/2\right) ^{2}}\ ,  \label{3eqstart}
\end{equation}%
%
%
%
%
%
%
%
%
%
%
%
%
%
%
%
%
%
%
%
%
%
%
%
where $n=1,2,3,\ldots $ are the quantum numbers, and the wave function in
terms of associated Laguerre polynomials is given by
\begin{equation}
\phi _{aa_{nl}}(\mathbf{r})=\tilde{\beta}\left[ \frac{(n-1-|l|)!}{\left[
(|l|+n-1)!\right] ^{3}(2n-1)}\right] ^{1/2}e^{-\tilde{\beta}r/2}(\tilde{\beta%
}r)^{|l|}L_{n+|l|-1}^{2|l|}(\tilde{\beta}r)\frac{e^{il\varphi }}{(2\pi
)^{1/2}}\ ,  \label{rk9}
\end{equation}%
where $\tilde{\beta}=e^{2}/\left[ (n-1/2)\varepsilon \mathcal{F}%
_{1}(\epsilon )\right] $, $\varphi $ is the polar angle, $L_{k}^{p}(x)$ are
the associated Laguerre polynomials, the quantum numbers $l$ can take the
values $l=0,\pm 1,\pm 2,\ldots ,\pm (n-1)$.


After simplification Eq.~(\ref{3eqstart})\ can be rewritten in the form of
the following quadratic equation
\begin{equation}
(C+8\gamma )\epsilon ^{2}-4\gamma k^{2}-4C\delta ^{2}=0\ ,  \label{eqstart2}
\end{equation}%
%
%
%
%
%
%
%
%
%
%
%
%
%
%
%
%
%
%
%
%
%
%
%
where $\gamma =(\hbar v_{F})^{2}$, $k=\hbar v_{F}\mathcal{K=}v_{F}P$, $%
C=e^{4}/\left( 4\pi \varepsilon _{0}\varepsilon (n-1/2)\right) ^{2}$.

The solutions of Eq.~(\ref{eqstart2}) are given by
\begin{equation}
\epsilon =2\left( \frac{\gamma (\hbar v_{F}\mathcal{K)}^{2}+C\delta ^{2}}{%
C+8\gamma }\right) ^{1/2}.  \label{eqn1}
\end{equation}%
%
%
%
%
%
%
%
%
%
%
%
%
%
%
%
%
%
%
%
%

Eq. (\ref{eqn1}) gives the energy-momentum dispersion $\epsilon (\mathcal{K}%
) $ of the electron and hole that are bound via Coulomb potential in a
single graphene layer. Since our interest is small kinetic energy,
therefore, for small $\mathcal{K}$ we expand Eq.~(\ref{eqn1}) with respect
to $\mathcal{K}^{2}$: and approximate $\epsilon (\mathcal{K})$ by the first
two terms of Taylor series
\begin{equation}
\epsilon =E_{b}+\frac{(\hbar \mathcal{K)}^{2}}{2\mathcal{M}}\ ,
\label{expand1}
\end{equation}%
where $E_{b}$ is the exciton binding energy is given by
\begin{equation}
E_{b}=2\delta \left( \frac{C}{C+2\gamma }\right) ^{1/2}\ ,  \label{binding}
\end{equation}%
and $\mathcal{M}$ is the effective mass of exciton given by
\begin{equation}
\mathcal{M}=\frac{\delta }{2\gamma v_{F}^{2}}\sqrt{(C+8\gamma )C}\ .
\end{equation}

\begin{figure}[ht]
\centering
\subfigure[]{
\includegraphics[width=0.48\textwidth]{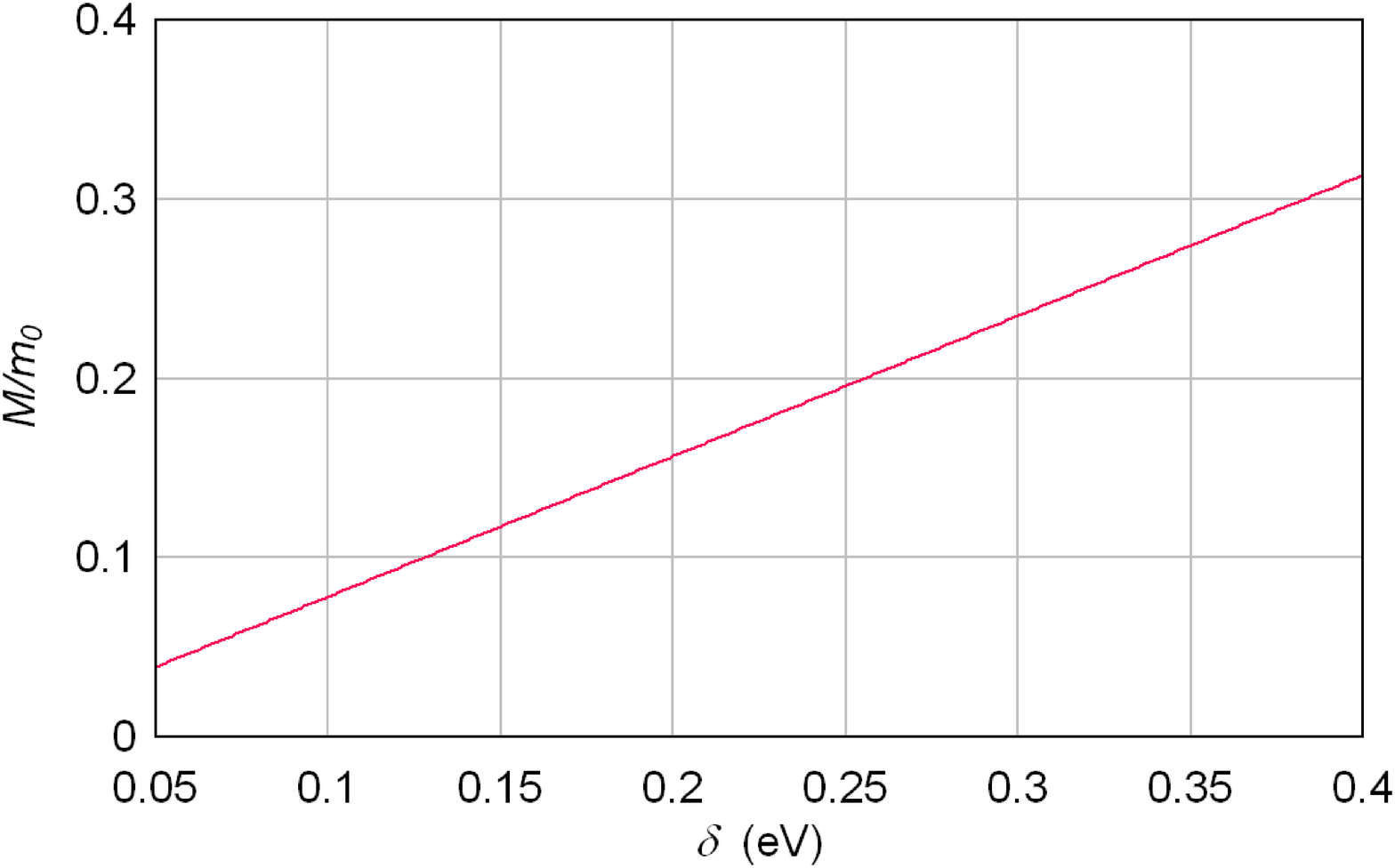}
} \subfigure[]{
\includegraphics[width=0.48\textwidth]{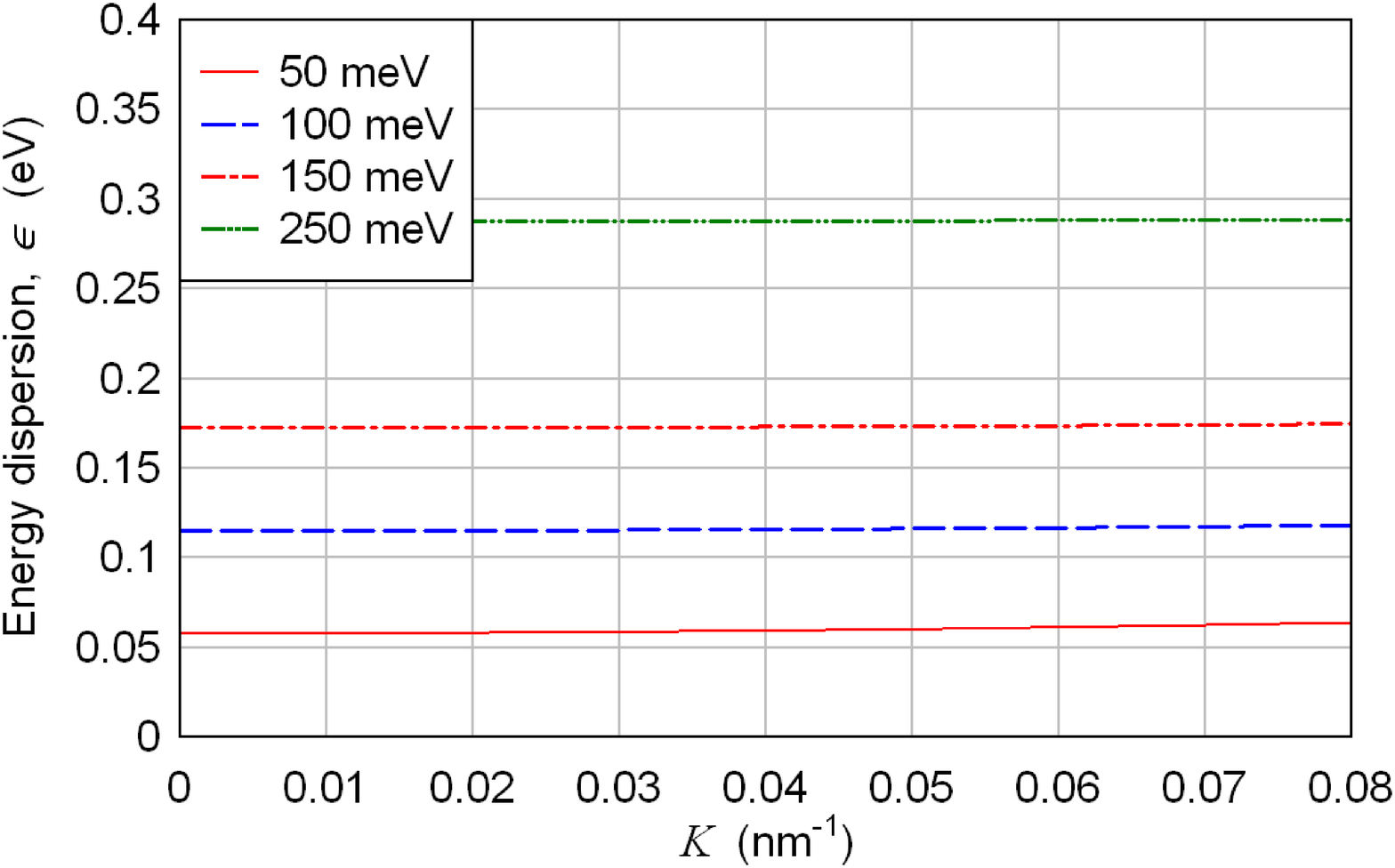}
}
\caption{Excitons in the single graphene layer. (a) The effective exciton
mass as a function of the energy gap.(b) The energy dispersion of exciton
for the different values of the energy gap.}
\label{F3}
\end{figure}

We note the exciton effective mass $\mathcal{M}$ increases when the gap $%
\delta $ increases as it is shown in Fig. \ref{F3}a. The result of
calculation of the exciton energy dispersion in graphene layer for different
gap energy $\delta $ in low energy parabolic approximation is given in Fig. %
\ref{F3}b. When $\mathcal{K\sim }$0.08 nm$^{-1}$ the parabolic approximation %
\eqref{expand1} gives less than 0.5\% difference with respect to the exact
solution \eqref{eqn1} and this value decreases when $\mathcal{K}$ deceases.
According to Fig.~\ref{F3}b, the exciton energy spectrum at the same quantum
number $n$ increases with the increase of the gap energy $\delta $. However,
for two graphene layers separated by the dielectric the energy dispersion
decreases when the total energy gap increases. Also the comparison of the
exciton energy distribution in graphene layer and in two graphene layers
separated by the dielectric shows that the exciton energy in graphene layer
always bigger than in two graphene layers separated by the dielectric.

\section{\protect\bigskip Conclusions}

\label{disc}

Now let us return to the fundamental and practical question related
to finding the solution of a problem of two interacting Dirac
particles that form the exciton in gapped graphene layer and in two
gapped graphene layers. In low-energy limit this problem can be
solved analytically, and we obtained the energy dispersion and wave
function of the exciton in gapped graphene layer and in two gapped
graphene layers separated by a dielectric. The excitons were
considered as a system of two Dirac particles interacting via a
Coulomb potential $V(r)$. In general case the center-of-mass and can
not be separated from the relative motion even though the
interaction depends only on the coordinate of the relative motion.
This is caused by the chiral nature of Dirac electron in graphene
and the electron and hole locations and momenta are 4-vectors and
must satisfy mass-shell conditions.\emph{\ }The analytical solution
for the energy dispersion and wave function was obtained by
introducing the transformation for the separation of the
center-of-mass and relative motions for two particles in graphene
that allows to reduce the Dirac-like equation for spinor to the
Schr\"{o}dinger-like second order differential equation for the
component of the spinor. In the parabolic approximation for the
energy dispersion found the effective mass of the exciton which are
the function of the energy gap in the single graphene layer and
function of the energy gaps and interlayer separation in the case of
two graphene layers separated by the dielectric. Firstly, we can
conclude that the exciton effective mass increases in the both cases
for a single graphene layer and for two layers graphene as the gap
energy increases. Also the exciton effective mass increases when the
interlayer separation increases. Therefore, by tuning the energy
gaps in graphene layers and changing interlayer separation one can
get the desirable value for the effective exciton mass. This is very
important for the system of many excitons when this system is
considered as a dilute gas of excitons that forms Bose-Einstein
condencate and undergoes to the Kosterlitz-Thouless phase transition
to a superfluid phase. By decreasing the mass of the exciton one can
increase the Kosterlitz-Thouless transition temperature. Secondly,
for the exciton in graphene layer the energy dispersion increases
with the increase of the gap energy. However, for the exciton in two
graphene layers separated by the dielectric the energy dispersion
decreases when the total energy gap increases as well as it
decreases when the interlayer separation increases.

\end{document}